\documentclass[11pt, oneside]{article}   	
\usepackage{geometry}                		
\usepackage{graphicx}				
\usepackage{amssymb}
\usepackage{amsmath}
\usepackage{braket}
\usepackage{mathtools}
\usepackage{indentfirst}
\usepackage{mathrsfs}
\usepackage{multicol}
\usepackage[numbers,sort&compress]{natbib}
\usepackage{MnSymbol}
\usepackage{xcolor}
\pagecolor{white}
\usepackage{overarrows}
\usepackage{setspace}
\usepackage{xcolor}

\usepackage{tgcursor}

\usepackage{booktabs}
\usepackage{multirow}
\usepackage{float}
\restylefloat{table}
\usepackage{siunitx} 
\usepackage{flafter}
\usepackage[format=hang,font=small,labelfont=bf]{caption}
\usepackage{float}
\usepackage{titlesec}
\usepackage{capt-of}
\usepackage{wrapfig}
\usepackage{placeins}
\usepackage{gensymb}

\newcommand{\Lagr}{\mathcal{L}}
\newcommand{\fock}{\textit{f}}
\newcommand{\nsum}{\displaystyle\sum_{I} ^{N^{\text{n}}}}
\newcommand{\csum}{\displaystyle\sum_{A}^{N^{\text{c}}}}
\newcommand{\nprod}{\displaystyle\prod_{I} ^{N^{\text{n}}}}
\newcommand{\esum}{\displaystyle\sum_{i} ^{N^{\text{e}}}}
\newcommand{\esumij}{\displaystyle\sum_{i>j} ^{N^{\text{e}}}}
\newcommand{\nsumij}{\displaystyle\sum_{I>J} ^{N^{\text{n}}}}
\newcommand{\lmhf}{{\boldsymbol{\mu}_I^{\gamma^{(0)}}}}
\newcommand{\lmmp}{{\boldsymbol{\mu}_I^{\gamma^{(2)}}}}
\newcommand{\lmvtwo}{{\boldsymbol{\mu}^\gamma}}
\newcommand{\lmnucone}{{\boldsymbol{\mu}_{n_2}^\gamma}}
\newcommand{\lmnuctwo}{{\boldsymbol{\mu}_{n_1}^\gamma}}
\newcommand{\hyll}{\mathcal{J}}

\usepackage{authblk}

\title{Constrained nuclear-electronic orbital \\ second-order M\o ller\textendash Plesset perturbation theory }
\author[1]{Gabrielle B. Tucker}
\author[1]{Kurt R. Brorsen\dag (brorsenk@missouri.edu)}
\affil[1]{Department of Chemistry, University of Missouri\textendash Columbia}

\begin{document}
\maketitle


\section{Abstract}
A multicomponent second-order M\o ller-Plesset perturbation theory (MP2) method is derived and implemented within the constrained nuclear-electronic orbital (CNEO) framework from a multicomponent generalization of the Hylleraas functional. The CNEO-MP2 method includes electronic-nuclear and nuclear correlation in the calculation of vibrationally averaged molecular properties. Nuclear quantum effects like vibrational averaging, isotopic effects, and zero-point energy can be captured in a single calculation or geometry optimization with CNEO-MP2, eliminating the need to perform costly subsequent calculations to determine higher order force constants as required with many existing methods used to determine vibrational effects upon molecular properties. The CNEO-MP2 method is benchmarked on a test set of diatomic and small polyatomic molecules and ions. Herein, we present internuclear distances, bond angles, potential energy surfaces, and vibrational frequencies calculated with the CNEO-MP2 method to demonstrate that it correctly captures the effects of nuclear vibrational motion upon molecular properties.

\section{Introduction}

Nuclear vibrations can have a significant impact upon molecular properties. \cite{ Bishop1990, Bettens, PuzzariniStanton}  For instance, vibrational contributions to rotational constants can range in magnitude from 0.1\% to 0.7\% of the size of the equilibrium rotational constant, which can lead to shifts of up to thousands of MHz.\cite{PuzzariniStanton} Vibrationally averaged $^{1}$H nuclear magnetic resonance spin-spin coupling constants can shift by up to nearly 40\% from their harmonic counterparts\cite{SolomonSchulman, Egidi}, and similarly, significant vibrational averaging effects have been observed for NMR chemical shifts\cite{GrigoleitBuhl, Tuttle, DracinskyHodgkinson}, chemical shift anisotropies and quadrupolar couplings\cite{Benzi}, dipolar interactions\cite{Ishii}, internuclear distances \cite{Toyama, Ishii}, isotropic hyperfine coupling constants\cite{Chen}, and optical rotation\cite{MortAutschbach}. While experimental measurements of molecular properties inherently include vibrational effects, molecular properties obtained from conventional electronic structure theory methods at the minimum-energy or equilibrium geometry of the Born-Oppenheimer potential energy surface (PES) under the harmonic approximation do not. 

Several methods have been introduced that can include the effects of nuclear vibrations  on geometries and molecular properties in electronic structure calculations including vibrational self-consistent field (VSCF) theory\cite{Bowman, Christiansen}, path integral approaches\cite{Shiga, GlaesemannFried, LopezCiudad}, and vibrational perturbation theory (VPT) \cite{Nielsen, Mills, NeugebauerHess, BloinoBarone, Franke, Fortenberry}. 

The most widely used of these methods is likely VPT.\cite{Franke} In VPT, the effects of anharmonic nuclear vibrations are treated as perturbations from a reference harmonic-oscillator Hamiltonian. In general, VPT correctly describes the effects of anharmonic nuclear vibration, including capturing differences upon isotopic substitution not observed in standard molecular property calculations. Despite this, VPT still has a number of significant limitations. Firstly, VPT can be computationally expensive as it commonly requires the calculation of third- or fourth-order force constants, which in many cases must be done numerically. Secondly, because the vibrational nuclear quantum effects are included after a standard electronic structure calculation that assumes the Born-Oppenheimer approximation, VPT does not treat the nuclei on equal footing with the electrons. Thirdly, perturbation theory works best for systems with small perturbations. Therefore, VPT is unlikely to perform well for systems with large vibrational anharmonicity or large amplitude motions.\cite{Franke} The goal of this work was thus to develop a method more computationally efficient than VPT, able to treat vibrational averaging and other nuclear quantum effects in a more comprehensive and direct manner, and able to work well for systems with significant vibrational anharmonicity. 

Alternative methods to VPT which can include vibrational effects in \textit{ab initio} electronic structure calculations are \textit{multicomponent} or \textit{pre-Born-Oppenheimer} methods. \cite{Matyus} In multicomponent methods, more than one type of particle (or component) is treated quantum mechanically by introducing a multicomponent Hamiltonian in which both electrons and (select) nuclei are treated on equal footing. \cite{NOMOone, NEOorig} This is in contrast to the standard (or single-component) methods of electronic structure theory where only the electrons are treated quantum mechanically. Due to the quantum treatment of the nuclei in multicomponent methods, these methods are capable of capturing a variety of nuclear quantum effects, as well as capturing nuclear isotopic effects in calculated molecular properties. \cite{Reyes, Khan, Smith}

The fundamental ideas of multicomponent methods were introduced by I. L. Thomas in the late 1960s. \cite{Thomas1, Thomas2} Since then, a variety of multicomponent frameworks have been developed with the most prominent likely being the nuclear-orbital molecular orbital (NOMO) approach of Tachikawa et al., \cite{NOMOone, Nakai2, Nakai} and the nuclear-electronic orbital (NEO) framework of the Hammes-Schiffer group. \cite{NEOorig, NEOrev} In recent years, the field of multicomponent methods has rapidly expanded and there now exist a variety of multicomponent methods for the treatment of electronic-nuclear correlation including  density functional theory (DFT) \cite{NEO96, NEO118, NEO105, Yang2017, Brorsen2017, Brorsen2018, NEO236, NEO255, Tao, NEO288, Hasecke, Gimferrer} and wavefunction-based \cite{NaSoHo, NEOorig, NakaiSodeyama, NEO57, NEO70, NEO75, Auer, NEO246, NEO259, Pav, Fajen, NEO298, Alaal, Fajen2, Fowler, Hasecke3, Hasecke2, Goudy, Fetherholf} methods. These methods have the same formal computational scaling (albeit with a larger prefactor) as their single-component counterparts. It should be noted that multicomponent methods are not limited to the treatment of electrons and nuclei, as positrons, muons, and other exotic particles can also be treated using multicomponent methods. \cite{Nieminen, Boronski, Makhov, Udagawa, Wiktor, Ellis, Rayka, Goli2, Rayka2, NEOrev, Goli, Riyahi, Pan} As the general approaches used in the different multicomponent frameworks are substantially similar, herein, we use the term NEO to describe these methods for simplicity. 

While NEO methods been shown to accurately include nuclear quantum effects neglected in standard single-component quantum chemistry methods, a historical limitation of the NEO framework is that it has required that at least two nuclei be treated classically to avoid contamination of the wavefunction by translational and rotational motion. \cite{TCON1, TCON2} While treating two or more nuclei classically removes any translational and rotational contamination, \cite{ATC} it comes at the cost of introducing a new Born-Oppenheimer-like separation between the quantum-mechanical and classical nuclei, which due to the much smaller mass difference between nuclei has less physical justification than the conventional Born-Oppenheimer separation between the electrons and the nuclei. \cite{NEO57, Yang2020b} As an alternative to treating at least two nuclei classically in multicomponent methods, there exist a variety of techniques developed in the broader class of pre-Born-Oppenheimer methods for the removal of translational and/or rotational contamination, although these techniques can be complicated to perform, especially for the removal of the rotational motion. \cite{Nakai2, TCON1, TCON2, HoshinoNakai, TRCON1}  

Recently, Xu and Yang introduced a modification to the standard NEO formalism, called the constrained nuclear-electronic orbital (CNEO) approach, that allows all nuclei in a system to be treated quantum mechanically. \cite{Yang2020a, Yang2020b} In the CNEO approach, it is assumed that the nuclei are still relatively well localized in comparison to the electrons. Therefore, constraints can be placed upon the position expectation values of each of the quantum nuclei in the relevant NEO Lagrangian. These constraints break the translational and rotational invariance of the NEO Lagrangian, which fixes the nuclear frame-of-reference, avoiding the need to treat any nuclei classically. In the CNEO approach, a PES-like surface is thus obtained as a function of the position expectation-values of the quantum nuclei as well as the coordinates of any classical nuclei. As discussed by Xu and Yang in the articles introducing the CNEO framework \cite{Yang2020a, Yang2020b}, this PES intrinsically includes nuclear quantum effects such as zero-point energy and is therefore suitable for the direct calculation of vibrationally averaged properties. Furthermore, with the CNEO approach nuclear quantum effects can be captured with a single calculation, potentially offering significant computational savings over most methods used to calculate vibrational corrections to single-component electronic properties. The CNEO approach has since been used for calculations on a wide variety of chemical systems where nuclear quantum effects are likely to be important, particularly in the contexts of molecular dynamics \cite{Zhao, Liu1, Liu2, Zhang} and spectroscopy \cite{Zhang2, Liu2, Xu, ChenHaoran}.
 
At present, the CNEO approach has been combined with multicomponent HF and DFT \cite{Yang2020a, Yang2020b}, as well as full configuration interaction (FCI) and unitary coupled cluster (UCC)\cite{Culpitt}. In this study, we introduce a new method for the calculation of vibrationally averaged molecular properties by combining the CNEO approach with multicomponent second-order M\o ller-Plesset perturbation theory (CNEO-MP2). 

This paper is organized as follows. In Section II, we present the theoretical background and show the derivation of the CNEO-MP2 method. In Section III, we discuss the implementation and the computational details related to the benchmarking of the new method. Then, in Section IV, we show that the CNEO-MP2 method performs well for describing the effects of the vibrational motion of nuclei upon molecular properties, capturing vibrational averaging effects and isotopic effects in addition to the zero point vibrational energy. Finally, we note that the theoretical formalism of the CNEO-MP2 method provides a framework for extending the CNEO approach to other many-body wavefunction methods such as multicomponent coupled-cluster theory. Therefore, the CNEO-MP2 method introduced in this study will provide a robust starting point for future developments.


\section{Theory}

\subsection{The NEO-HF Method}
In the NEO formalism\cite{NEOorig}, the Hamiltonian for a system of \textit{N}$^{\text{e}}$ electrons, \textit{N}$^{\text{c}}$ classical nuclei, and \textit{N}$^{\text{n}}$ quantum nuclei, is 
\begin{equation}
\begin{aligned}
\hat{H}_{\text{NEO}} = & - \esum\frac{\nabla^{2}_{i}}{2} - \esum\csum\frac{Z_{A}}{|\boldsymbol{r}^{\text{e}}_{i} - \boldsymbol{R}^{\text{c}}_{A}|}  + \esumij\frac{1}{|\boldsymbol{r}^{\text{e}}_{i} - \boldsymbol{r}^{\text{e}}_{j}|} \\
& - \nsum\frac{\nabla^{2}_{I}}{2m_{I}} + \nsum\csum\frac{Z_{A}Z_{I}}{|\boldsymbol{R}^{\text{n}}_{I} - \boldsymbol{R}^{\text{c}}_{A}|} + \nsumij\frac{Z_{I}Z_{J}}{|\boldsymbol{R}^{\text{n}}_{I} - \boldsymbol{R}^{\text{n}}_{J}|} \\
& - \nsum\esum\frac{Z_{I}}{|\boldsymbol{r}^{\text{e}}_{i} - \boldsymbol{R}^{\text{n}}_{I}|},
\end{aligned}
\end{equation}
where lowercase variables and indices are used for the electrons, and uppercase variables and indices are used for the nuclei.
The NEO-HF wavefunction ansatz for the electrons and quantum nuclei is a direct product of an electronic Slater determinant and a product over nuclear determinants or permanents depending upon the fermionic or bosonic nature of the quantum nuclei:
\begin{equation}
 \ket{\Psi} =  \ket{\Phi^{\text{e}}} \otimes (\displaystyle\prod_{t} \ket{\Phi^{\text{n}_{t}}}).
\end{equation}
Here, \textit{t} indexes the type of the nucleus, and $\text{n}_{t}$ indicates a nucleus of type \textit{t}. 
At the NEO Hartree-Fock level of theory, a constraint term ensuring the orthonormality of the nuclear molecular orbitals is included along with the familiar electronic molecular orbital orthonormality constraint in the Lagrangian expression for optimization of the NEO-HF energy:
\begin{equation}
\Lagr = E_{\text{NEO-HF}}  -  \nsum \epsilon^{\text{n}}_{I}  (\braket{\phi^{\text{n}}_{I} | \phi^{\text{n}}_{I}} - 1)   -  \esum \epsilon^{\text{e}}_{i}  (\braket{\phi^{\text{e}}_{i} | \phi^{\text{e}}_{i}} - 1).
\end{equation}
Here, \textit{E}$_{\text{NEO-HF}}$ is the NEO-HF energy, $\phi^{\text{e}}_{i}$ and $\phi^{\text{n}}_{I}$ are the electronic and nuclear molecular orbitals, respectively, and $\epsilon^{\text{e}}_{i}$ and $\epsilon^{\text{n}}_{I}$ are the Lagrange multipliers for the electronic and nuclear constraint terms, respectively.
The molecular orbitals which characterize the determinants and or permanents of the NEO-HF wavefunction are approximated as linear combinations of atomic orbitals, shown for the electronic orbitals in equation 4, 
\begin{equation}
\phi_{i}^{\text{e}} = \sum_{r} C_{ri}^{\text{e}}\chi_{r}^{\text{e}},
\end{equation}
and for the nuclear orbitals in equation 5,
\begin{equation}
 \phi_{I}^{\text{n}} = \sum_{R} C_{RI}^{\text{n}}\chi_{R}^{\text{n}}.
\end{equation}
Minimization of the NEO-HF energy with respect to variation of the molecular orbitals under the electronic and nuclear orthonormality constraints leads to a set of Roothaan-Hall-like matrix equations for the electrons, 
\begin{equation}
\textbf{F}^{\text{e}}\textbf{C}^{\text{e}} = \textbf{S}^{\text{e}}\textbf{C}^{\text{e}}\textbf{E}^{\text{e}},
\end{equation}
and for the nuclei,
\begin{equation}
\textbf{F}^{\text{n}}\textbf{C}^{\text{n}} = \textbf{S}^{\text{n}}\textbf{C}^{\text{n}}\textbf{E}^{\text{n}},
\end{equation}
respectively. We note that there is a Roothann-Hall-like matrix equation for each type of nuclei, but show only one for simplicity. The exact form of the nuclear Fock matrices can be found in previous studies. \cite{NOMOone, NEOorig, Nakai}

\subsection{The NEO-MP2 Method}
The second-order M\o ller-Plesset perturbation theory (MP2) method has previously been derived and implemented for the treatment of correlation in the multicomponent NEO framework.\cite{NakaiSodeyama, NEO70} In the NEO-MP2 method, the total energy, \textit{E}$_{\text{NEO-MP2}}$, is a sum of the NEO-HF energy with the second-order electronic, nuclear, and electronic-nuclear correlation energies:
\begin{equation}
E_{\text{NEO-MP2}} = E_{\text{NEO-HF}} + E^{(2)}_{\text{ee}} + E^{(2)}_{\text{nn}} + E^{(2)}_{\text{en}},
\end{equation}
where the second-order electronic correlation energy is defined as
\begin{equation}
E^{(2)}_{\text{ee}} = \frac{1}{4}\displaystyle\sum_{ijab}\frac{|\braket{ij|ab} - \braket{ij|ba}|^{2}}{\epsilon^{\text{e}}_{i} + \epsilon^{\text{e}}_{j} - \epsilon^{\text{e}}_{a} - \epsilon^{\text{e}}_{b}},
\end{equation} 
the second-order electronic-nuclear correlation energy is defined as
\begin{equation}
E^{(2)}_{\text{en}} = \displaystyle\sum_{M}^{N^{\text{n}}}\{\displaystyle\sum_{iIaA}\frac{|\braket{iI|aA}|^{2}}{\epsilon^{\text{e}}_{i} + \epsilon^{\text{n}}_{I} - \epsilon^{\text{e}}_{a} - \epsilon^{\text{n}}_{A}}\}.
\end{equation}
and the second-order nuclear correlation energy is defined as
\begin{equation}
E^{(2)}_{\text{nn}} = \displaystyle\sum_{MN}\{\frac{1}{4}\displaystyle\sum_{\substack{IA \in \{\phi^{\text{n}}_{M}\} \\JB \in \{\phi^{\text{n}}_{N}\}}}\frac{|\braket{IJ|AB} - \delta_{MN}\braket{IJ|BA}|^{2}}{\epsilon^{\text{n}}_{I} + \epsilon^{\text{n}}_{J} - \epsilon^{\text{n}}_{A} - \epsilon^{\text{n}}_{B}}\}.
\end{equation}
Throughout this work, the electronic indices for generic orbitals will be denoted by \textit{p}, \textit{q}, \textit{r}, \textit{s}, ...,  indices for occupied orbitals by \textit{i}, \textit{j}, \textit{k}, \textit{l}, ..., and indices for virtual orbitals by \textit{a}, \textit{b}, \textit{c}, \textit{d}, ..., with the analogous capital indices used for the nuclear orbitals. The indices \textit{M} and \textit{N} are used to sum over nuclei, with orbitals \textit{I} and \textit{A} associated with nucleus \textit{M}, and \textit{J} and \textit{B} associated with nucleus \textit{N}. Note that as there is no exchange between particles of different kinds, the numerator of the electronic-nuclear correlation energy includes only the Coulomb term. Similarly, for any contributions to the correlation energy involving pairs of nuclei of different types there are no exchange terms for these pairs as indicated by the Kronecker delta in the numerator of the nuclear correlation energy.

The NEO-MP2 correlation energy can also be determined from the multicomponent coupled-cluster equations\cite{Pav}, or by minimizing the multicomponent Hylleraas functional with respect to the first order wavefunction amplitudes\cite{Fajen}.

\subsection{The CNEO-HF Method}
With the nuclei being significantly more massive than the electrons, they are expected to remain relatively localized in comparison to the electrons for most chemical systems, resulting in little overlap between nuclear orbitals. Following from this assumption, in the CNEO approach, nuclei of the same type are treated as distinguishable particles. This constitutes making the the Hartree Product approximation for the nuclear wavefunction.\cite{Auer} Thus, the CNEO wavefunction, originally a direct product of an electronic Slater determinant wavefunction with nuclear Slater determinant or permanent wavefunctions depending on the fermionic or bosonic nature of the nuclei, reduces to a direct product of the electronic Slater determinant with a simple Hartree product over the nuclear orbitals. 

\begin{equation}
 \ket{\Psi} =  \ket{\Phi^{\text{e}}} \otimes (\nprod \ket{\Phi_{I}^{\text{n}}}).
\end{equation}

By making the Hartree product approximation for the nuclear wavefunction, nuclear exchange effects are neglected in CNEO methods, although these have been previously shown to be negligible relative to the other two-particle contributions for calculations performed with NEO-HF and NEO-DFT on the water hexamer and glycine. \cite{Auer} Furthermore, the use of this approximation has been demonstrated to provide significant computational savings through a reduction in the total number of two-particle integrals which must be calculated, as well as a reduction in the size of the nuclear matrices requiring diagonalization during the self-consistent field (SCF) procedure. \cite{Auer} 

In the CNEO-HF approach, the nuclear orbitals are constrained such that the expectation value for the nuclear position at the Hartree-Fock level is equal to the position specified for each nucleus:
\begin{equation}  
\nsum \lmhf \cdot (\braket{\phi^{\text{n}}_{I} | \hat{r} | \phi^{\text{n}}_{I}} - {\textbf{R}^{\text{n}}_I})  = 0, 
\end{equation} where the $\boldsymbol{\mu}_{I}^{\gamma^{(0)}}$ are the Lagrange multipliers for the density constraints.
Because the expectation value of the nuclear density is required to remain fixed at the chosen nuclear position coordinates, all nuclei in a molecular system may be treated quantum mechanically without introducing translational or rotational contamination. When all nuclei are treated quantum mechanically, the CNEO PES becomes a function of the expectation values of the positions of all quantum nuclei and inherently incorporates nuclear quantum effects. \cite{Yang2020a, Yang2020b}

The Lagrangian expression for optimization of the CNEO-HF energy includes the new position expectation value constraint terms in addition to the orthonormality constraints on the electronic and nuclear orbitals, and is defined as:
\begin{equation}
\begin{aligned}
\Lagr[\phi^{\text{e}}, \phi^{\text{n}}] = E_{\text{CNEO-HF}}  +  & \nsum \lmhf \cdot (\braket{\phi^{\text{n}}_{I} | \hat{r} | \phi^{\text{n}}_{I}} -\textbf{R}^{\text{n}}_{I}) -  \nsum \epsilon^{\text{n}}_{I}  (\braket{\phi^{\text{n}}_{I} | \phi^{\text{n}}_{I}} - 1)   -  \esum \epsilon^{\text{e}}_{i}  (\braket{\phi^{\text{e}}_{i} | \phi^{\text{e}}_{i}} - 1),   
\end{aligned}
\end{equation} 
where \textit{E}$_{\text{CNEO-HF}}$ is the CNEO-HF energy. Minimizing this Lagrangian expression leads to a modified set of multicomponent Roothaan-Hall-like matrix equations which are solved for the electronic and nuclear orbitals, respectively. \cite{Yang2020a, Yang2020b}

\subsection{The CNEO-MP2 Method}

\subsubsection{Multicomponent Hylleraas Correlation Energies}
Herein, we combine the existing CNEO-HF and NEO-MP2 frameworks. In the same way that the MP2 energy can be derived from the Hylleraas functional,
\begin{equation}
\hyll_{2} [\textbf{t}] = 2 \Re (\braket{\Psi | \hat{V} | \Psi_{0}}) + \braket{\Psi | \hat{H_{0}} - E_{n}^{(0)} | \Psi},
\end{equation}
the NEO-MP2 energy can be derived from a multicomponent form of the Hylleraas functional, such that
\begin{equation}
\hyll_{2} [\textbf{t}] = \hyll_{2} [\textbf{t}^{\text{e}}, \textbf{t}^{\text{en}}, \textbf{t}^{\text{n}}] = \hyll_{\text{e}}[\textbf{t}^{\text{e}}] + \hyll_{\text{en}}[\textbf{t}^{\text{en}}] + \hyll_{\text{n}}[\textbf{t}^{\text{n}}],
\end{equation}
where $\textbf{t}^{\text{e}}$, $\textbf{t}^{\text{en}}$, and $\textbf{t}^{\text{n}}$ are the electronic, electronic-nuclear, and nuclear t-amplitudes, respectively.
Following the derivation of multicomponent orbital-optimized MP2 of Pavo\v{s}evi\'{c} et al. for the coupled cluster t-amplitudes and $\Lambda$-equations within the conventional NEO framework \cite{Pav}, and adapting the nuclear and electronic-nuclear expressions to treat all nuclei, the multicomponent NEO-MP2 Hylleraas expressions are:
\begin{equation}
 \hyll_{\text{e}}[\textbf{t}^{\text{e}}] = \displaystyle\sum_{ijab} \{ \frac{1}{2} \displaystyle\sum_{c} (\fock_{ac} \lambda_{ij}^{ab} \textit{t}_{cb}^{ij}) - \frac{1}{2} \displaystyle\sum_{k} (\fock_{ik} \lambda_{kj}^{ab} \textit{t}_{ab}^{ij}) + \frac{1}{4}(\overline{\textit{g}}_{ij}^{ab}\textit{t}_{ab}^{ij}) + \frac{1}{4}(\overline{\textit{g}}_{ab}^{ij}\lambda_{ij}^{ab})\},
  \end{equation}

\begin{equation}
 \hyll_{\text{en}}[\textbf{t}^{\text{en}}] = \displaystyle\sum_{M}^{N^n} \{ \displaystyle\sum_{iIaA} \{ \displaystyle\sum_{c}(\fock_{ac} \lambda_{iI}^{aA} \textit{t}_{cA}^{iI})  - \displaystyle\sum_{k} (\fock_{ik}\lambda_{kI}^{aA}\textit{t}_{aA}^{iI}) + \displaystyle\sum_{C} (\fock_{AC}\lambda_{iI}^{aA}\textit{t}_{aC}^{iI}) - \displaystyle\sum_{K} (\fock_{IK}\lambda_{iK}^{aA}\textit{t}_{aA}^{iI}) - (\textit{g}_{iI}^{aA}\textit{t}_{aA}^{iI}) - (\textit{g}_{aA}^{iI}\lambda_{iI}^{aA}) \} \},
 \end{equation}
and 
\begin{equation}
\begin{aligned}
 \hyll_{\text{n}}[\textbf{t}^{\text{n}}] = & \displaystyle\sum_{MN} \{ \displaystyle\sum_{\substack{IA \in \{\phi^{\text{n}}_{M}\}\\JB \in\{\phi^{\text{n}}_{N}\}}} \{ \displaystyle\sum_{C_{M}}(\fock_{AC_{M}} \lambda_{IJ}^{AB} \textit{t}_{C_{M}B}^{IJ}) +\displaystyle\sum_{C_{N}}(\fock_{BC_{N}} \lambda_{IJ}^{AB} \textit{t}_{AC_{N}}^{IJ}) \\
                                 & - \displaystyle\sum_{K_{M}} (\fock_{IK_{M}} \lambda_{K_{M}J}^{AB} \textit{t}_{AB}^{IJ}) - \displaystyle\sum_{K_{N}} (\fock_{JK_{N}} \lambda_{IK_{N}}^{AB} \textit{t}_{AB}^{IJ}) \\
                                 & + (\textit{g}_{IJ}^{AB}\textit{t}_{AB}^{IJ}) + (\textit{g}_{AB}^{IJ}\lambda_{IJ}^{AB}) \}\}, 
\end{aligned}                                 
\end{equation} respectively. The indices $M$ and $N$ are again used to distinguish between different nuclei in the sums. The lowercase $f$ here represent elements of the NEO Fock matrix, the lowercase $g$ represent antisymmetrized two-particle integrals for pairs of electrons or simply coulomb integrals for mixed electron-nucleus pairs or for pair of nuclei, the lowercase $t$ represent the t-amplitudes, and the lowercase $\lambda$ represent the $\Lambda$-amplitudes.

\subsubsection{Expectation Value Constraint on the MP2 Single-Particle Nuclear Density}
At the MP2 level, the single-particle nuclear density matrix for nucleus \textit{n}$_{M}$, $\gamma^{(2)n_{M}}$, is given by the expressions:
\begin{equation}  
\gamma_{AB}^{(2)n_{M}} [\textbf{t}^{\text{en}}, \textbf{t}^{\text{n}}]  = \lambda_{IJ}^{AC} \textit{t}_{BC}^{IJ} + \lambda_{iI}^{aA} \textit{t}_{aB}^{iI}
\end{equation}
for the virtual block, and 
\begin{equation}  
\gamma_{IJ}^{(2)n_{M}} [\textbf{t}^{\text{en}}, \textbf{t}^{\text{n}}]  = - \lambda_{JK}^{AB} \textit{t}_{AB}^{IK} - \lambda_{iJ}^{aA} \textit{t}_{aA}^{iI}
\end{equation}
for the occupied block, respectively.
We introduce an additional constraint at the level of the MP2 single-particle nuclear density, such that the Lagrangian expression for CNEO-MP2 becomes:
\begin{equation}
\Lagr [\textbf{t}^{\text{e}}, \textbf{t}^{\text{en}}, \textbf{t}^{\text{n}}] = \hyll_{\text{e}}[\textbf{t}^{\text{e}}] + \hyll_{\text{en}}[\textbf{t}^{\text{en}}] + \hyll_{\text{n}}[\textbf{t}^{\text{n}}] + \nsum \displaystyle\sum_{PQ} \lmmp \cdot (\gamma^{(2)n_{I}}_{PQ} [\textbf{t}^{\text{en}}, \textbf{t}^{\text{n}}] \braket{\phi^{\text{n}_{I}}_{P} | \hat{r} | \phi^{\text{n}_{I}}_{Q}} - \textbf{R}^{\text{n}}_{I}),
\end{equation}
where $\boldsymbol{\mu}_{I}^{\gamma^{(2)}}$ is used for the Lagrange multipliers of the correlated nuclear density constraints. 

\subsubsection{Non-canonical MP2}

To obtain the CNEO-HF energy, as well as optimized molecular orbital coefficients which satisfy the imposed nuclear density constraints which are needed to construct the Fock matrix, we first perform a CNEO-HF calculation. To ensure well-defined orbital energies, necessary for performing the MP2 calculations, we use the conventional NEO expressions to build the core Hamiltonian matrix and construct the Fock matrices, 
\begin{equation}
\begin{aligned}
f^{\text{e}}_{\mu\nu} = & h^{\text{e}}_{\mu\nu} +  \displaystyle\sum_{i} ^{N_{\text{occ.}}^{\text{e}}} \displaystyle\sum_{\lambda\sigma} (C_{\lambda i}C^{*}_{\sigma i}) \{ 2\braket{\mu\nu | \lambda\sigma} - \braket{\mu\nu | \sigma\lambda}\},
\end{aligned}
\end{equation}  
and
\begin{equation}
\begin{aligned}
f^{\text{n}}_{\mu\nu} = & h^{\text{n}}_{\mu\nu} + \displaystyle\sum_{I} ^{N_{\text{occ.}}^{\text{n}}} \displaystyle\sum_{\lambda\sigma} (C_{\lambda I}C^{*}_{\sigma I}) \{ \braket{\mu\nu | \lambda\sigma} \}.
\end{aligned}
\end{equation}
However, by using the CNEO-HF molecular orbital coefficients to construct the conventional NEO-HF Fock matrix, the nuclear Fock matrices are no longer diagonal and thus are non-canonical, which requires the use of non-canonical or iterative MP2 for the CNEO-MP2 method. Here, in the atomic orbital basis the lowercase $\mu$, $\nu$, $\lambda$, and $\sigma$ are used for both electrons and nuclei to avoid notational complications.  We note that definitions of the Fock operator other than that of NEO-HF could be used for CNEO-MP2 as well as for other CNEO many-body methods. Other possibilities include using the CNEO-HF Fock operator which would result in a canonical CNEO-MP2 Fock matrix or a semi-canonical Fock operator where only the occupied-occupied and virtual-virtual blocks of the CNEO-MP2 Fock matrix are diagonal, but the exploration of the resultant effects of using such Fock operators are beyond the scope of the present study.

To calculate the CNEO-MP2 energy correction, in addition to the elements of the non-canonical CNEO-MP2 Fock matrix, both the multicomponent MP2 t-amplitudes and $\Lambda$-amplitudes are required. Note that here the $\Lambda$-amplitudes are simply the complex conjugates of the t-amplitudes.

The electronic, electronic-nuclear, and nuclear residual expressions are obtained by differentiating the CNEO-MP2 Lagrangian expression with respect to the electronic, electronic-nuclear, and nuclear t-amplitudes, and are given by the expressions:

\begin{equation}
\begin{aligned}
R_{ab}^{ij} = & \displaystyle\sum_{ijab}\{\braket{ij || ab}  + \displaystyle\sum_{c} ^{N^{\text{vir.}}} (\fock_{ac} \textit{t}_{cb}^{ij} + \fock_{bc} \textit{t}_{ac}^{ij}) - \displaystyle\sum_{k} ^{N^{\text{occ.}}} (\fock_{ki} \textit{t}_{ab}^{kj} + \fock_{kj} \textit{t}_{ab}^{ik})\}, 
\end{aligned}
\end{equation}

\begin{equation}
\begin{aligned}
R_{aA}^{iI} = &\displaystyle\sum_{iIaA}\{ -\braket{iI | aA} + \displaystyle\sum_{c} ^{N^{\text{vir.}}} (\fock_{ac} \textit{t}_{cA}^{iI}) + \displaystyle\sum_{C} ^{N^{\text{vir.}}} (\fock_{AC} \textit{t}_{aC}^{iI}) 
 - \displaystyle\sum_{k} ^{N^{\text{occ.}}} (\fock_{ki} \textit{t}_{aA}^{kI}) - \displaystyle\sum_{K} ^{N^{\text{occ.}}} (\fock_{KI} \textit{t}_{aA}^{iK}) \\
&  + \displaystyle\sum_{C}^{N^{\text{vir.}}} \lmvtwo \cdot (\lambda_{iI}^{aC} + t_{aC}^{iI}) \braket{\phi^{n}_{C} | \hat{r} | \phi^{n}_{I}} +\displaystyle\sum_{K}^{N^{\text{occ.}}} \lmvtwo \cdot (\lambda_{iK}^{aA} + t_{aA}^{iK}) \braket{\phi^{n}_{K} | \hat{r} | \phi^{n}_{I}}\},
\end{aligned}
\end{equation}
and
\begin{equation}
\begin{aligned}
R_{AB}^{IJ} = & \displaystyle\sum_{IJAB}\{\braket{IJ | AB}  + \displaystyle\sum_{C} ^{N^{\text{vir.}}} (\fock_{AC} \textit{t}_{CB}^{IJ} + \fock_{BC} \textit{t}_{AC}^{IJ}) - \displaystyle\sum_{K} ^{N^{\text{occ.}}} (\fock_{KI} \textit{t}_{AB}^{KJ} + \fock_{KJ} \textit{t}_{AB}^{IK}) \\
& +  \displaystyle\sum_{C}^{N^{\text{vir.}}_{\text{n}_{1}}} \lmnuctwo \cdot (\lambda_{IJ}^{CB} + t_{CB}^{IJ}) \braket{\phi^{\text{n}_{1}}_{C} | \hat{r} | \phi^{\text{n}_{1}}_{A}} + \displaystyle\sum_{K}^{N^{\text{occ.}}_{\text{n}_{1}}} \lmnuctwo \cdot (\lambda_{KJ}^{AB} + t_{AB}^{KJ}) \braket{\phi^{\text{n}_{1}}_{K} | \hat{r} | \phi^{\text{n}_{1}}_{I}} \\
& +  \displaystyle\sum_{C}^{N^{\text{vir.}}_{\text{n}_{2}}} \lmnucone \cdot (\lambda_{IJ}^{AC} + t_{AC}^{IJ}) \braket{\phi^{\text{n}_{2}}_{C} | \hat{r} | \phi^{\text{n}_{2}}_{B}} + \displaystyle\sum_{K}^{N^{\text{occ.}}_{\text{n}_{2}}} \lmnucone \cdot (\lambda_{IK}^{AB} + t_{AB}^{IK}) \braket{\phi^{\text{n}_{2}}_{K} | \hat{r} | \phi^{\text{n}_{2}}_{J}}\}.
\end{aligned}
\end{equation}
These multicomponent residual expressions are used to solve for the electronic, electronic-nuclear, and nuclear t-amplitudes, respectively. It is important to note that the last two terms in the electronic-nuclear residual, and the last four terms in the nuclear MP2 residual, arise due to the constraint introduced for the correlated nuclear density in the Lagrangian expression for optimizing the multicomponent Hylleraas correlation energies. Upon rearrangement one obtains the following expressions:

\begin{equation}
\begin{aligned}
(\epsilon_{i} + \epsilon_{j} - \epsilon_{a} - \epsilon_{b}) \textit{t}_{ab}^{ij} = & \displaystyle\sum_{ijab}\{ \braket{ij || ab}  \\
& + \displaystyle\sum_{c \neq a} ^{N^{\text{vir.}}} (\fock_{ac} \textit{t}_{cb}^{ij}) + \displaystyle\sum_{c \neq b} ^{N^{\text{vir.}}} (\fock_{bc} \textit{t}_{ac}^{ij}) - \displaystyle\sum_{k \neq i} ^{N^{\text{occ.}}} (\fock_{ki} \textit{t}_{ab}^{kj}) - \displaystyle\sum_{k \neq j} ^{N^{\text{occ.}}}(\fock_{kj} \textit{t}_{ab}^{ik})\} \\
\end{aligned}
\end{equation}
for the electronic t-amplitudes,
\begin{equation}
\begin{aligned}
(\epsilon_{i} + \epsilon_{I} - \epsilon_{a} - \epsilon_{A}) \textit{t}_{aA}^{iI} = &\displaystyle\sum_{iIaA}\{ -\braket{iI | aA} \\
&+ \displaystyle\sum_{c \neq a} ^{N^{\text{vir.}}} (\fock_{ac} \textit{t}_{cA}^{iI}) + \displaystyle\sum_{C \neq A} ^{N^{\text{vir.}}} (\fock_{AC} \textit{t}_{aC}^{iI}) - \displaystyle\sum_{k \neq i} ^{N^{\text{occ.}}} (\fock_{ki} \textit{t}_{aA}^{kI}) - \displaystyle\sum_{K \neq I} ^{N^{\text{occ.}}} (\fock_{KI} \textit{t}_{aA}^{iK}) \\
&  \displaystyle\sum_{C}^{N^{\text{vir.}}} \lmvtwo \cdot (\lambda_{iI}^{aC} + t_{aC}^{iI}) \braket{\phi^{\text{n}}_{C} | \hat{r} | \phi^{\text{n}}_{A}} +\displaystyle\sum_{K}^{N^{\text{occ.}}} \lmvtwo \cdot (\lambda_{iK}^{aA} + t_{aA}^{iK}) \braket{\phi^{\text{n}}_{K} | \hat{r} | \phi^{\text{n}}_{I}}\}\end{aligned}
\end{equation}
for the electronic-nuclear t-amplitudes, and
\begin{equation}
\begin{aligned}
(\epsilon_{I} + \epsilon_{J} - \epsilon_{A} - \epsilon_{B}) \textit{t}_{AB}^{IJ} = & \displaystyle\sum_{IJAB}\{ \braket{IJ | AB}  \\
& + \displaystyle\sum_{C \neq A} ^{N^{\text{vir.}}} (\fock_{AC} \textit{t}_{CB}^{IJ}) + \displaystyle\sum_{C \neq B} ^{N^{\text{vir.}}} (\fock_{BC} \textit{t}_{AC}^{IJ})  - \displaystyle\sum_{K \neq I} ^{N^{\text{occ.}}} (\fock_{KI} \textit{t}_{AB}^{KJ}) - \displaystyle\sum_{K \neq J} ^{N^{\text{occ.}}}(\fock_{KJ} \textit{t}_{AB}^{IK}) \\
& +  \displaystyle\sum_{C}^{N^{\text{vir.}}_{\text{n}_{1}}} \lmnuctwo \cdot (\lambda_{IJ}^{CB} + t_{CB}^{IJ}) \braket{\phi^{\text{n}_{1}}_{C} | \hat{r} | \phi^{\text{n}_{1}}_{A}} + \displaystyle\sum_{K}^{N^{\text{occ.}}_{\text{n}_{1}}} \lmnuctwo \cdot (\lambda_{KJ}^{AB} + t_{AB}^{KJ}) \braket{\phi^{\text{n}_1}_{K} | \hat{r} | \phi^{\text{n}_1}_{I}} \\
& +  \displaystyle\sum_{C}^{N^{\text{vir.}}_{\text{n}_{2}}} \lmnucone \cdot (\lambda_{IJ}^{AC} + t_{AC}^{IJ}) \braket{\phi^{\text{n}_{2}}_{C} | \hat{r} | \phi^{\text{n}_{2}}_{B}} + \displaystyle\sum_{K}^{N^{\text{occ.}}_{\text{n}_{2}}} \lmnucone \cdot (\lambda_{IK}^{AB} + t_{AB}^{IK}) \braket{\phi^{\text{n}_{2}}_{K} | \hat{r} | \phi^{\text{n}_{2}}_{J}}\}
\end{aligned}
\end{equation}
for the nuclear t-amplitudes, each of which can be divided through by the sum of the occupied and virtual orbital energies (the diagonal matrix elements of the non-canonical CNEO-MP2 Fock matrix) to isolate the t-amplitudes of interest.
The use of non-canonical Fock matrices requires the MP2 t-amplitudes to be solved for iteratively, as the  the off-diagonal Fock matrix terms in each summation are no longer guaranteed to vanish. 
For the iterative procedure, we require a solution of both the t-amplitudes and the Lagrange multipliers. 

The minimization of the residual is performed for the electronic, and electronic-nuclear, and nuclear t-amplitudes at each iteration step, and the constraint expressions for the nuclear and electronic-nuclear amplitudes can be satisfied using a root finding algorithm. This is executed in a double loop structure, similarly to that used in the CNEO-DFT study which extended the constraint upon the nuclear density to all nuclei in a system.\cite{Yang2020b} First, we make initial guesses for the Lagrange multipliers and each of the sets of t-amplitudes. In the outer loop in each iteration, a new guess is constructed for the t-amplitudes using the Lagrange multipliers optimized for the previous iteration for construction of the electronic-nuclear and nuclear amplitudes. In the inner loop, the Lagrange multipliers are optimized for the sets of electronic-nuclear and nuclear t-amplitudes constructed in the outer loop. The determination of the nuclear and electronic-nuclear t-amplitudes was accomplished using the Levenberg-Marquardt algorithm\cite{Levenberg, Girard, Wynne, Morrison, Marquardt} as implemented in the SciPy\cite{SciPy} optimize module for root finding. In both the inner and outer loop structures, the Hylleraas CNEO-MP2 correlation energies and root-mean-square differences (RMSD) in the t-amplitudes are tested for convergence. Additionally, Pulay's direct inversion in the iterative subspace (DIIS) technique \cite{Pulay} is used to accelerate convergence of the t-amplitudes in both loops. 

\section{Methods}

\subsection{Implementation}
The CNEO-MP2 method was implemented in a locally modified version of a fork of PySCF that contains the CNEO-HF and CNEO-DFT methods and is maintained by the Yang group. \cite{PySCF1, PySCF2, YangGithub} Performance critical functions were written using Cython\cite{Cython}. The CNEO-MP2 code is available at https://github.com/ brorsenk/CNEO-MP2. 

\subsection{Computational Details}
We use several benchmarks to demonstrate the performance of the CNEO-MP2 method. First, for a test set of diatomic and small polyatomic molecules and ions we performed geometry optimizations to determine vibrationally averaged bond lengths. Second, we looked at the potential energy landscapes associated with the bifluoride and deuterium bifluoride anions. Third, we compare proton affinities calculated with single-component MP2 with those calculated with multicomponent NEO-MP2 and CNEO-MP2 to investigate the performance of the new method in calculating thermochemical properties. Lastly, we investigate the vibrational frequency of the shared proton mode in a simple water cluster, the Zundel cation (H$_{5}$O$_{2}^{+}$). We also note that calculated nuclear densities, including on-axis and off-axis slices, are presented in the Supplemental Material.

All CNEO-MP2 calculations used the PB4-D nuclear basis set \cite{PB4D} for hydrogen, scaled PB4-D for deuterium, and a 12s12p12d even-tempered nuclear basis set\cite{Ruedenberg} with the parameters $\alpha=2\sqrt{2}m$ and $\beta=\sqrt{3}$ for all other nuclei. For the geometry optimizatons the cc-pVDZ, cc-pVTZ, cc-pVQZ, aug-cc-pVDZ, aug-cc-pVTZ, and aug-cc-pVQZ electronic basis sets were used. \cite{Dunning}  Recently, there has been interest in the development of new electronic basis sets designed particularly for multicomponent calculations, however, we chose here to use standard electronic basis sets to mirror the calculations in the original CNEO studies. Additionally, for many calculations in this study, we treat all nuclei quantum mechanically whereas many of the newly introduced electronic basis sets are only available for hydrogen.\cite{Samsonova, FabijanBasis, HammesSchifferBasis} 

Geometry updates were performed using the Nelder-Mead optimization algorithm \cite{NM} as implemented in the SciPy module \cite{SciPy} for the diatomic and triatomic molecules, and using the Broyden-Fletcher-Goldfarb-Shanno (BFGS) optimization algorithm \cite{Broyden, Fletcher, Goldfarb, Shanno, Greenstadt} as implemented in SciPy for all other systems. For the diatomic and triatomic systems, all nuclei were treated quantum mechanically, and all calculations were performed with the uncorrelated density constrained, and both with and without the correlated density constrained. For the four-atom molecules, calculations were performed with only the hydrogen nuclei treated quantum mechanically, and with both the uncorrelated and correlated nuclear densities constrained. For the Zundel cation, only the shared proton was treated quantum mechanically, and the geometry optimization was performed with only the uncorrelated density constrained. For comparison, geometry optimizations were also performed for all systems with single-component second-order M\o ller-Plesset perturbation theory (MP2) using ORCA (v. 5.0.3) \cite{ORCA1, ORCA2} and Gaussian (v.16) \cite{Gaussian}, with an additional VPT2-MP2 calculation performed in Gaussian at the single-component MP2 minimum subsequent to the geometry optimzation. 

Single-component MP2 and multicomponent CNEO-MP2 potential energy surfaces (PESs) were calculated and plotted for the bifluoride anion ($\textrm{FHF}^{\textrm{-}}$) and the deuterium bifluoride anion  ($\textrm{FDF}^{\textrm{-}}$). The single-component MP2 PES was calculated with the aug-cc-pVTZ electronic basis set, while the CNEO-MP2 PES was calculated with the aug-cc-pVTZ electronic basis set, the PB4-D nuclear basis set for hydrogen, a scaled PB4-D nuclear basis set for deuterium, and the 12s12p12d nuclear basis set for fluorine, with all nuclei treated quantum mechanically. For both calculations, the molecule was aligned on the z-axis, and the energy of the system is plotted as a function of the displacement of the hydrogen atom from the origin in the xz-plane. The fluorine atoms were fixed at -1.145 {\AA} and 1.145 {\AA}, respectively. Plots of the surfaces were generated from the position and energy data using the Matplotlib module in Python \cite{matplotlib}. 

Proton affinities were calculated for a test set of 11 small molecules and ions with single-component MP2, multicomponent NEO-MP2, and multicomponent CNEO-MP2. The geometries for all systems in the test set were optimized in ORCA (v. 6.0.1) with the aug-cc-pVQZ electronic basis set, with the frozen-core approximation (used by default for MP2 calculations in ORCA) turned off. The multicomponent NEO-MP2 and CNEO-MP2 single-point energy calculations used to determine the multicomponent proton affinites were performed at these geometries optimized with single-component MP2. For all of the multicomponent calculations, the aug-cc-pVQZ electronic basis set was used, and only the added proton was treated quantum mechanically using the PB4-D nuclear basis set.

The vibrational frequency for the shared proton mode of the Zundel cation (H$_{5}$O$_{2}^{+}$) was calculated with the CNEO-MP2 method, aug-cc-pVTZ electronic basis set, and PB4-D nuclear basis set. Numerical Hessian calculations were performed on the optimized geometry, both without and with the constraint upon the correlated density (in both cases the uncorrelated density was constrained). For comparison, a single-component MP2 numerical Hessian was performed in ORCA (v. 6.0.1) to obtain the vibrational frequency of the shared proton mode.

We finally note that, as mentioned above, we have examined the effects of not including the constraint upon the correlated density by performing many of the benchmarking calculations both with the constraint, which will be denoted CCD for constrained correlated density, and without the constraint, which will be denoted UCD for unconstrained correlated density. For the UCD calculations, the same residual equations presented in the Theory section above are solved, but without including the terms that depend upon the Lagrange multipliers. Therefore, the loop to solve for the Lagrange multipliers becomes unnecessary, and only a single loop is required to solve for the t-amplitudes, a benefit of which is a reduction in the computational cost of the CNEO-MP2 method. As shown in the Supplemental Materials, we observe no significant shift in the correlated CNEO-MP2 nuclear densities relative to those of CNEO-HF for the FHF$^{-}$ and HCN molecules and no differences were found between the bond lengths calculated with or without the constraint included for the correlated density.  The agreement between the CCD and UCD results is somewhat expected for CNEO-MP2, as it is known that NEO-MP2 gives nuclear densities almost identical to NEO-HF. However, we stress that the agreement between the CCD and UCD results will likely not be the case for more advanced post-HF CNEO methods such as CNEO-CC that we plan to develop in the future. Therefore, we choose to use and show results using the CCD. The full CNEO-MP2 CCD and UCD results are included in the Supplemental Material.


\section{Results and Discussion}
\subsection{Geometry Analysis}
The calculated internuclear distances for three isotopes of molecular hydrogen are displayed in Table 1. Because the effect of the nuclear masses on internuclear distance is not accounted for in the single-component MP2 method, there are no differences among the isotopologues. For the VPT2-MP2 results,  as the nuclear masses are accounted for in the calculations, the deuterium-substituted isotopes have shorter internuclear distances. Furthermore, for the VPT2-MP2 results, the calculated internuclear distances are in each case longer than those calculated with single-component MP2. This behavior is expected as internuclear distances should increase upon averaging over vibrational quantum states when anharmonicity is present in the potential. 

The CNEO-MP2 results, like the VPT2-MP2 results, exhibit the correct behavior of the internuclear distances with respect to isotopic substitution (shorter bond lengths for more deuterated isotopes), as well as with respect to increase in the distances upon the inclusion of vibrational effects.  

\begin{table}[H]
\begin{center}
\caption{Internuclear distances calculated for the isotopes of molecular hydrogen (\AA ngstroms). Results are presented for the single-component MP2 method, the single-component VPT2-MP2 method, and the multicomponent CNEO-MP2 method. Calculations were performed using the aug-cc-pVDZ (aDZ), aug-cc-pVTZ (aTZ), and aug-cc-pVQZ (aQZ) electronic basis sets.}
\begin{tabular}{|| c c | c c c ||} 
\hline
&  & MP2 & VPT2-MP2 & CNEO-MP2 \\
\hline
H$_{2}$ & aDZ & 0.755 & 0.778 & 0.775 \\
 & aTZ & 0.737 & 0.761 & 0.765 \\
 & aQZ & 0.736 & 0.760 & 0.762 \\
HD & aDZ & 0.755 & 0.775 & 0.772  \\
 & aTZ & 0.737 & 0.758 & 0.761 \\
 & aQZ & 0.736 & 0.757 & 0.759 \\
 D$_{2}$ & aDZ & 0.755 &  0.771 &  0.769 \\ 
 & aTZ & 0.737 & 0.754 & 0.757 \\
 & aQZ & 0.736 & 0.753 & 0.755  \\
\hline
\end{tabular}
\end{center}
\end{table}

When comparing internuclear distances calculated with the different electronic basis sets, a general trend for all systems is a significant decrease in the distances calculated with the cc-pVTZ and aug-cc-pVTZ basis sets compared to those calculated with the cc-pVDZ and aug-cc-pVDZ basis sets. For the distances calculated with the cc-pVQZ and aug-cc-pVQZ basis sets, there was no discernable trend when comparing to the cc-pVTZ and aug-cc-pVTZ basis set distances. For simplicity, for all other molecular systems benchmarked, we present only the aug-cc-pVQZ results as these are expected to be the most accurate.

The full sets of internuclear distances calculated with the cc-pVDZ, cc-pVTZ, cc-pVQZ, aug-cc-pVDZ, aug-cc-pVTZ, and aug-cc-pVQZ electronic basis sets for all molecules and methods are included in the Supplemental Materials. 

The aug-cc-pVQZ internuclear distances calculated for four more diatomics, HF, DF, OH$^{-}$, and OD$^{-}$, are presented in Table 2, those calculated for the triatomics HCN,  DCN, HNC, DNC, FHF$^{-}$, FDF$^{-}$, H$_{2}$O, HDO, and D$_{2}$O are presented in Table 3, and  those calculated for H$_{2}$CO and H$_{2}$O$_{2}$ are presented in Table 4. For these diatomic and triatomic systems, all nuclei were treated quantum mechanically for the multicomponent methods whereas for the four-atom systems only the hydrogen atoms were treated quantum mechanically for computational efficiency. It should be noted that for the four-atom systems, in some cases there was a slight variation in some of the bond angles of the same type and in one case for the bond lengths for the H$_{2}$O$_{2}$ results, which could likely be resolved by adjusting the criteria for convergence used in the CNEO-MP2 calculations (here we used a convergence tolerance of 10$^{-8}$ for the energies and RMSD densities). In general, the same trends held for all of these systems that were discussed above with regard to the electronic basis sets, the increase in internuclear distances upon vibrational averaging, and the isotopic effects for the deuterium substituted species.

\begin{table}[H]
\begin{center}
\caption{Internuclear distances for HF, DF, OH$^{-}$, and OD$^{-}$ calculated with the aug-cc-pVQZ electronic basis  (\AA ngstroms). Results are presented for the single-component MP2, single-component VPT2-MP2, and CNEO-MP2 CCD methods. For the CNEO-MP2 calculations, all nuclei were treated quantum mechanically with the PB4-D nuclear basis set used for hydrogen, scaled PB4-D for deuterium, and 12s12p12d for all heavier atoms.}
\begin{tabular}{|| c | c c c ||}
\hline
 & MP2 & VPT2-MP2 & CNEO-MP2 (CCD) \\
\hline
HF & 0.919 & 0.934 & 0.931 \\
DF &  0.919 &  0.930 &  0.926 \\ 
OH$^{-}$ & 0.965 & 0.981 & 0.976 \\
OD$^{-}$ &  0.965 & 0.976 &  0.972 \\
\hline
\end{tabular}
\end{center}
\end{table}

\begin{table}[H]
\begin{center}
\caption{Internuclear distances for HCN, DCN, HNC, DNC, H$_{2}$O, HDO, D$_{2}$O, FHF$^{-}$, and FDF$^{-}$ calculated with the aug-cc-pVQZ electronic basis set (\AA ngstroms). Bond angles in degrees are also included for the water isotopologues. Results are presented for the single-component MP2, single-component VPT2-MP2, and multicomponent CNEO-MP2 CCD methods, with some comparisons to ground vibrational state effective geometries obtained from experiment when available. For the CNEO-MP2 CCD calculations, all nuclei were treated quantum mechanically with the PB4-D nuclear basis set used for hydrogen, scaled PB4-D for deuterium, and 12s12p12d for all heavier atoms.}

\begin{tabular}{|| c | c c c||}
\hline
 & MP2 & VPT2-MP2 & CNEO-MP2 \\
\hline
HCN: H\textendash C & 1.065 & 1.099 & 1.075 \\
HCN: C\textendash N & 1.164 & 1.178 & 1.157 \\
DCN: D\textendash C & 1.065 & 1.100 & 1.070 \\
DCN: C\textendash N & 1.164 & 1.179 & 1.157 \\
HNC: H\textendash N & 0.996 & 1.031 & 1.006 \\
HNC: N\textendash C & 1.174 & 1.192 & 1.167 \\
DNC: D\textendash N & 0.996 & 1.032 & 1.002 \\
DNC: N\textendash C & 1.174 & 1.192 & 1.166 \\
FHF$^{-}$: H\textendash F & 1.143 & 1.111 & 1.148 \\
FDF$^{-}$: D\textendash F & 1.143 & 1.109 & 1.146 \\
H$_{2}$O: O\textendash H & 0.959 & 0.973 & 0.971 \\
HDO: O\textendash H & 0.959 & 0.972 & 0.971 \\
HDO: O\textendash D & 0.959 & 0.970 & 0.966 \\
D$_{2}$O: O\textendash D & 0.959 & 0.969 & 0.966 \\
H$_{2}$O: $\theta$(H\textendash O\textendash H) & 104.270 & 104.167 & 104.953 \\
HDO: $\theta$(H\textendash O\textendash D) & 104.270 & 104.175 & 104.939 \\
D$_{2}$O: $\theta$(D\textendash O\textendash D) & 104.270 & 104.169 & 104.923 \\

\hline
\end{tabular}
\end{center}
\end{table}

\begin{table}[H]
\begin{center}
\caption{Internuclear distances (\AA ngstroms) and bond angles (degrees) for H$_{2}$CO and H$_{2}$O$_{2}$ calculated with the aug-cc-pVQZ electronic basis set. Results are presented for the single-component MP2, VPT2-MP2, and multicomponent CNEO-MP2 CCD methods. For the multicomponent CNEO-MP2 CCD calculations, only hydrogen nuclei were treated quantum mechanically with the PB4-D nuclear basis set.}

\begin{tabular}{|| c | c c c ||}
\hline 
& MP2 & VPT2-MP2 & CNEO-MP2 \\
\hline
H$_{2}$CO: C\textendash O & 1.209 & 1.214 & 1.207 \\
H$_{2}$CO: C\textendash H & 1.099 & 1.112 & 1.118 \\
H$_{2}$CO: $\theta$ (O\textendash C\textendash H) & 121.714 & 121.683 & 121.653 \\
H$_{2}$CO: $\theta$ (H\textendash C\textendash H) & 116.572 & 116.635 & 116.695 \\
H$_{2}$O$_{2}$: H(2)\textendash O(0) & 0.964 & 0.965 &  0.981 \\
H$_{2}$O$_{2}$: O(0)\textendash O(1) & 1.447 & 1.461 & 1.447 \\
H$_{2}$O$_{2}$: O(1)\textendash H(3) & 0.964 & 0.965 & 0.981 \\
H$_{2}$O$_{2}$: $\theta$ (H(2)\textendash O(0)\textendash O(1))& 99.800 & 99.608 & 99.910 \\
H$_{2}$O$_{2}$: $\theta$ (H(3)\textendash O(1)\textendash O(0))& 99.800 & 99.608 & 99.920 \\
H$_{2}$O$_{2}$: $\tau$ (H(2)\textendash O(0)\textendash O(1) \textendash H(3))& 112.665 & 116.520 & 113.898\\
\hline
\end{tabular}
\end{center}
\end{table}
We note again that when anharmonicity is present in the vibrational potential of molecules, the average geometry of the ground vibrational state is generally longer than that of the equilibrium geometry found at the minimum of the vibrational potential well. Geometries optimized to minima with standard Born-Oppenheimer quantum chemistry calculations are harmonic equilibrium geometries. Geometries optimized with CNEO methods, however, include anharmonic vibrational effects and thus will resemble more closely the experimentally measured average geometries of ground vibrational states rather than equilibrium molecular geometries. 

To better put the computational results for the molecular geometries investigated into context, we look at two systems from our test set in more detail for which experimental ground state vibrational geometries were available in the literature, HCN and H$_{2}$CO. The equilibrium geometry of HCN for instance has an H\textendash C bond of 1.063 {\AA} and a C\textendash N bond of 1.154 {\AA}, whereas the average ground vibrational state geometry has an H\textendash C bond of 1.064 {\AA} and a C\textendash N bond of 1.156 \AA.\cite{hcn_ref} Similarly for H$_{2}$CO, the equilibrium geometry has an H\textendash C bond of 1.100 {\AA} and a C\textendash O bond of 1.203 {\AA} \cite{formaldehyde_eq}, whereas the average ground vibrational state geometry has an H\textendash C bond of 1.117 {\AA} and a C\textendash O bond of 1.206 {\AA} \cite{formaldehyde_ro}. For HCN, the MP2 results closely reproduce the H\textendash C equilibrium and vibrational ground state bond lengths, but overestimate the C\textendash N equilibrium and vibrational ground state bond lengths by 0.09 to 0.07 {\AA} respectively. The VPT2-MP2 results shift towards the vibrational ground state geometry from the equilibrium values for both bonds but are significantly overestimated. The CNEO-MP2 results shift the H\textendash C bond lengths in the correct direction, but overestimate by 0.011 {\AA}, whereas for the C\textendash bond length, not only does it shift in the correct direction, it is within 0.001 {\AA}  of the vibrational ground state value. For H$_{2}$CO, the MP2 results are in relatively good agreement with the equilibrium geometry, the VPT2-MP2 bonds lengths shift toward the ground vibrational state bond lengths underestimating by 0.005 {\AA} for the C\textendash H bond length and overestimating by 0.008 {\AA} for the C\textendash O bond, and the CNEO-MP2 bond lengths most closely reproduce the vibrational ground state bond lengths, overestimating the C\textendash H bond length by only 0.001 {\AA} and overestimating the C\textendash O bond length by 0.001 {\AA}.

We emphasize that not only does the CNEO-MP2 method correctly capture the changes due to nuclear quantum effects on molecular geometries, it is able to do so in a single geometry optimization calculation, which offers the potential for significant computational savings relative to VPT2, as VPT2 requires the additional calculation of third order and sometimes higher order force constants post geometry optimization.

\subsection{Proton Affinities}
Proton affinities were calculated for N$_{2}$, OH$^{-}$, SH$^{-}$, CN$^{-}$, CO$_{2}$, H$_{2}$O, H$_{2}$S, NO$_{2}^{-}$, CH$_{2}$O, NH$_{3}$, and HCOO$^{-}$, with single-component MP2, multicomponent NEO-MP2, and multicomponent CNEO-MP2, respectively, and are presented in Table 5 in eV. For the single-component MP2 proton affinities, the geometries of all systems were optimized in ORCA 6.0.1 using the aug-cc-pVQZ electronic basis set and with the frozen core approximation turned off, and numerical hessians were calculated for each system to determine the zero-point energy contributions. The MP2 proton affinities were calculated using the final energies of the optimized structures and include the zero-point energy contributions. The equations used to calculate the single-component and multicomponent proton affinities can be found in previous NEO-DFT and NEO-CC studies of multicomponent proton affinities. \cite{Brorsen2017, Pav} 

The geometries optimized for each system with MP2 in ORCA 6.0.1 were subsequently used to perform single-point calculations to determine the energies with NEO-MP2, and CNEO-MP2 in PySCF for determination of the multicomponent proton affinities. For the multicomponent NEO-MP2 and CNEO-MP2 proton affinities, the zero-point energy of the quantum proton is inherently incorporated, however the zero-point energies for the modes associated with the classical nuclei were not accounted for. The single-component MP2 method performed most accurately in the calculation of proton affinities, with a mean absolute error of 0.24 eV. The mean absolute error for the multicomponent NEO-MP2 and CNEO-MP2 methods was slightly higher than that of the single-component MP2 error, at 0.53 eV and 0.54 eV, respectively. For MP2, the maximum error was 1.00 eV which occured for H$_{2}$CO, while for both NEO-MP2 and CNEO-MP2 the maximum errors of 0.59 eV occurred for H$_{2}$S. It is not unexpected that the MP2, NEO-MP2 and CNEO-MP2 methods are not sufficiently accurate for the calculation of energetic properties due to the insufficient treatment of correlation in second-order perturbation theory, though for CNEO-MP2 it is possible that the error may be improved through performing calculations with all nuclei treated quantum mechanically, which would provide a full treatment of the zero-point energy contributions. Previously reported values for NEO-MP2 proton affinities for these systems reported smaller mean absolute errors, however these values may be artificially lowered due to the mixed basis set treatment used for the electronic basis sets (where aug-cc-pVQZ was used with the quantum proton and aug-cc-pVTZ with all other atoms), which can take advantage of error cancellation between the basis set error and the correlation error. \cite{Pav}

\newpage
\begin{table}[H]
\begin{center}
\caption{Proton affinities for a test set of diatomic and small polyatomic molecules and ions were calculated with single-component MP2, and multicomponent NEO-MP2 and CNEO-MP2. Geometry optimizations for each system were performed with single-component MP2. Single point calculations at the MP2 optimized geometries were performed with NEO-MP2 and CNEO-MP2 to calculate the multicomponent proton affinities. All calculations used the aug-cc-pVQZ electronic basis set. For the NEO-MP2 and CNEO-MP2 calculations, only the added hydrogen nucleus was treated quantum mechanically with the PB4-D nuclear basis set. \\ $^{\alpha}$Experimental proton affinities used for comparison were obtained from References \cite{proaff1, proaff2, proaff3, proaff4, proaff5, proaff6}.}
\begin{tabular}{|| c | c c c c ||}
\hline
SYSTEM & EXPT.$^{\alpha}$ & MP2 & NEO-MP2 & CNEO-MP2\\
\hline
N$_{2}$ & 4.93 & 5.03 & 4.71 & 4.70 \\
OH$^{-}$ & 16.93 & 17.37 & 16.39 & 16.39 \\
SH$^{-}$ & 15.31 & 15.31 & 14.72 & 14.72 \\
CN$^{-}$ & 15.31 & 15.18 & 14.79 & 14.78 \\
CO$_{2}$ & 5.60 & 5.72 & 5.15 & 5.14 \\
H$_{2}$O & 7.16 & 7.32 & 6.76 & 6.75 \\
H$_{2}$S & 7.31 & 7.42 & 6.90 & 6.90 \\
NO$_{2}^{-}$ & 14.75 & 15.00 & 14.20 & 14.20 \\
H$_{2}$CO & 7.39 & 6.39 & 6.96 & 6.96 \\
NH$_{3}$ & 8.85 & 9.01 & 8.48 & 8.47 \\
HCOO$^{-}$ & 14.97 & 15.22 & 14.46 & 14.46 \\
Mean AE & \textendash & 0.24 & 0.45 & 0.46 \\
Max AE & \textendash & 1.00 & 0.59 & 0.59 \\

\hline
\end{tabular}
\end{center}
\end{table}

\subsection{Potential Energy Surfaces}
PESs for the bifluoride and deuterium bifluoride anions were calculate using a grid-based approach. With the molecules oriented along the z-axis, and the fluorine atoms fixed at 1.146 and -1.146 \AA, respectively, a grid of 900 points was used to vary the geometry of the central hydrogen or deuterium atoms. The y-coordinate was held fixed, and the region from -0.4 to 0.4 \AA{} on the z-axis (the molecular axis), and from -0.8 to 0.8 \AA{} on the x-axis was investigated. PESs were generated for each of the two molecules using single-component MP2 and multicomponent CNEO-MP2. All calculations used the aug-cc-pVTZ electronic basis set. The CNEO-MP2 calculations used the PB4-D basis set for hydrogen, a scaled PB4-D basis set for deuterium, and a 12s12p12d even-tempered nuclear basis set for fluorine. The single-component FHF$^{-}$, single-component FDF$^{-}$, multicomponent FHF$^{-}$, and multicomponent FDF$^{-}$ PESs are presented in Figures 1-4, respectively. In comparing the single-component PESs for FHF$^{-}$,and FDF$^{-}$ it can be seen that, as expected, there is no difference between them. When looking at the multicomponent PESs, it can be seen that the depths of the wells are shifted more positively than those of the single-component surfaces due to the inherent inclusion of the zero-point vibrational energy in the CNEO-MP2 method, similarly to those calculated with single-component DFT and CNEO-DFT presented in the study which introduced the CNEO framework.\cite{Yang2020a} The central parts of the single-component potential wells are more oblong in character, while the multicomponent potential wells including the vibrational effects are more broad with a somewhat more round character in the central area. Comparing the multicomponent FHF$^{-}$ and FDF$^{-}$ PESs to one another, FDF$^{-}$ has a deeper well, and the energies are shifted more negatively than those of  FHF$^{-}$. The regions of the multicomponent FDF$^{-}$ PES below -198.240 Hartree are also more broad than those of the FHF$^{-}$ PES below this energy (this encompasses much of the central region of the well between roughly -0.2 and 0.2 \AA{} along the z-axis and between nearly -0.25 and 0.25 \AA{} along the x-axis for FDF$^{-}$, and roughly -0.1 and 0.1 \AA{} along the z-axis and -0.1 to 0.1 \AA{} along the x-axis for FHF$^{-}$. The broader depth of the central part of the well for FDF$^{-}$ is expected as FDF$^{-}$ has a shorter bond length than FHF$^{-}$ due to the mass of the deuterium, as well as due to the fact that during the exploration of the PESs the fluorine atoms were fixed at the total F\textendash F distance optimized for FHF$^{-}$ with the aug-cc-pVTZ electronic basis set during the benchmarking of the bond lengths for these molecules. It is clear from the differences in the shape, depth, and character of the multicomponent potential wells of the isotopologues from one another that the CNEO-MP2 method is able to capture the expected isotopic effects upon the molecular properties.

\begin{figure}
                    \centering
                    \includegraphics[width=8.5cm]{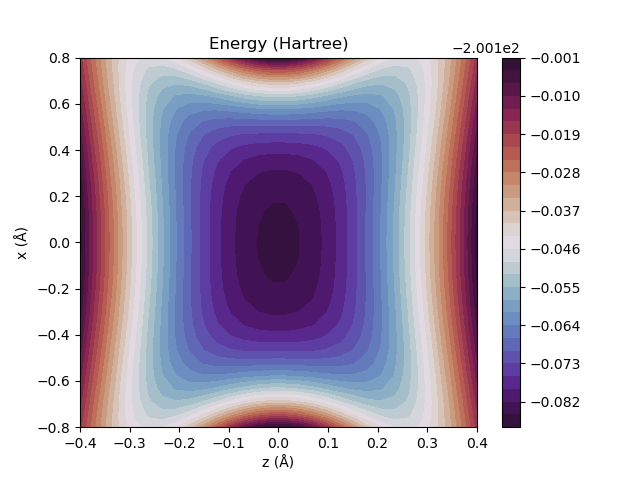}
                                \caption{Single-component MP2 potential energy surface of the bifluoride anion (FHF$^{-}$) as a function of the position of the hydrogen atom in the xz-plane. The fluorine atoms are fixed at 1.146 and -1.146 \AA.}
        \label{fig:placeholder}
    \end{figure}

\begin{figure}
    \centering
    \includegraphics[width=8.5cm]{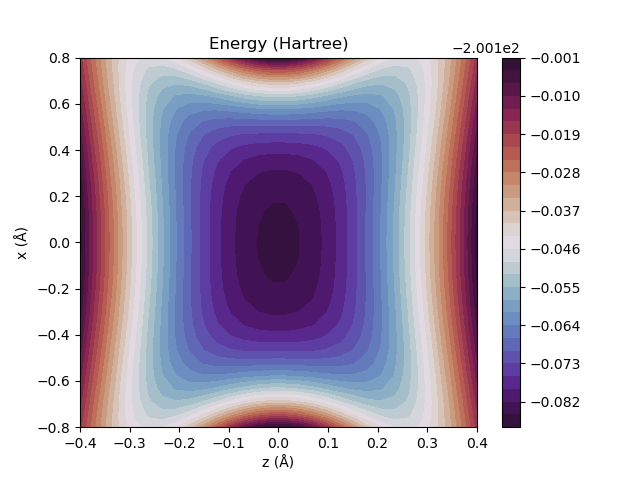}
        \caption{Single-component MP2 potential energy surface of the deuterium bifluoride anion (FDF$^{-}$) as a function of the position of the deuterium atom in the xz-plane. The fluorine atoms are fixed at 1.146 and -1.146 \AA.}
    \label{fig:placeholder}
\end{figure}

\begin{figure}
    \centering
    \includegraphics[width=8.5cm]{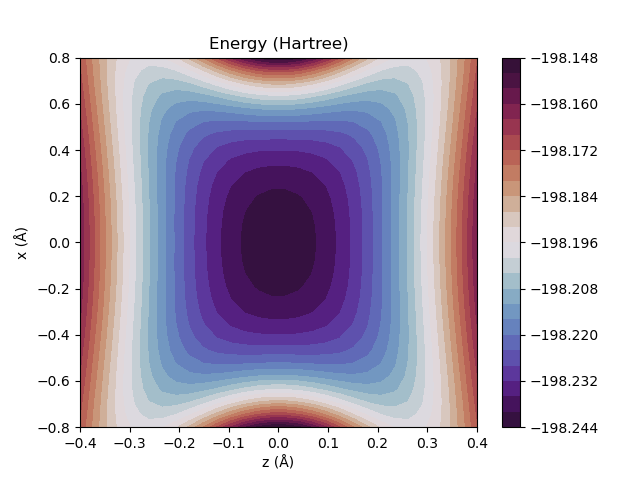}
        \caption{Multicomponent CNEO-MP2 potential energy surface of the bifluoride anion (FHF$^{-}$) as a function of the position of the hydrogen atom in the xz-plane. Here the fluorine atoms are fixed at 1.146 and -1.146 \AA.}
    \label{fig:placeholder}
\end{figure}

\begin{figure}
            \centering
            \includegraphics[width=8.5cm]{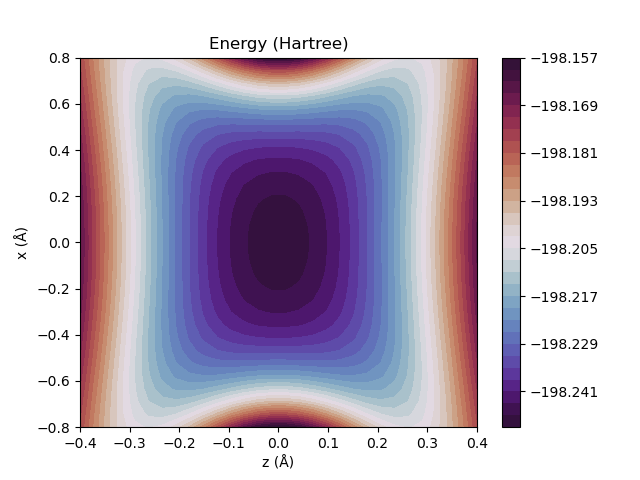}
                \caption{Multicomponent CNEO-MP2 potential energy surface of the deuterium bifluoride anion (FDF$^{-}$) as a function of the position of the deuterium atom in the xz-plane. Here the fluorine atoms are fixed at 1.146 and -1.146 \AA.}
    \label{fig:placeholder}
\end{figure}

\FloatBarrier

\subsection{The Zundel Cation: Geometry and Frequencies}
The Zundel cation (H$_{5}$O$_{2}^{+}$), is an important system for studying proton transfer and other effects of nuclear vibration due to the relatively flat and highly anharmonic nature of the potential energy surface associated with the shared proton mode.\cite{Xie, Huang, YangZundelA} As such, we investigated the performance of the CNEO-MP2 method for calculating the geometry and vibrational frequencies of the Zundel cation. Geometry optimizations followed by numerical Hessian calculations were performed with single-component MP2 in ORCA (v. 6.0.1), with VPT2-MP2 in Gaussian 16, and with CNEO-MP2 in PySCF. For both the single-component and multicomponent calculations, the aug-cc-pVTZ electronic basis set was used. The multicomponent CNEO-MP2 calculations used the PB4-D basis set with only the shared proton treated quantum mechanically to reduce the computational cost. Previous multicomponent calculations for this system with CNEO-DFT have used the same basis set, however in these all hydrogen atoms were treated quantum mechanically. \cite{YangZundelA} For the single-component calculations performed in ORCA, the frozen-core approximation is used by default for the MP2 gradient calculations. For the CNEO-MP2 calculations, the geometry was optimized without the constraint upon the correlated density using the BFGS optimization algorithm. The optimized structures were all confirmed to be minima from the numerical Hessian calculations, and all have C$_{2}$ symmetry, except for the VPT2-MP2 structure, which was assigned to the C$_{1}$ symmetry group. The CNEO-MP2 optimized structure is displayed in Figure 5, and the bond lengths and angles are presented in Table 6. 
\newpage
\begin{figure}
    \centering
    \includegraphics[width=8.5cm]{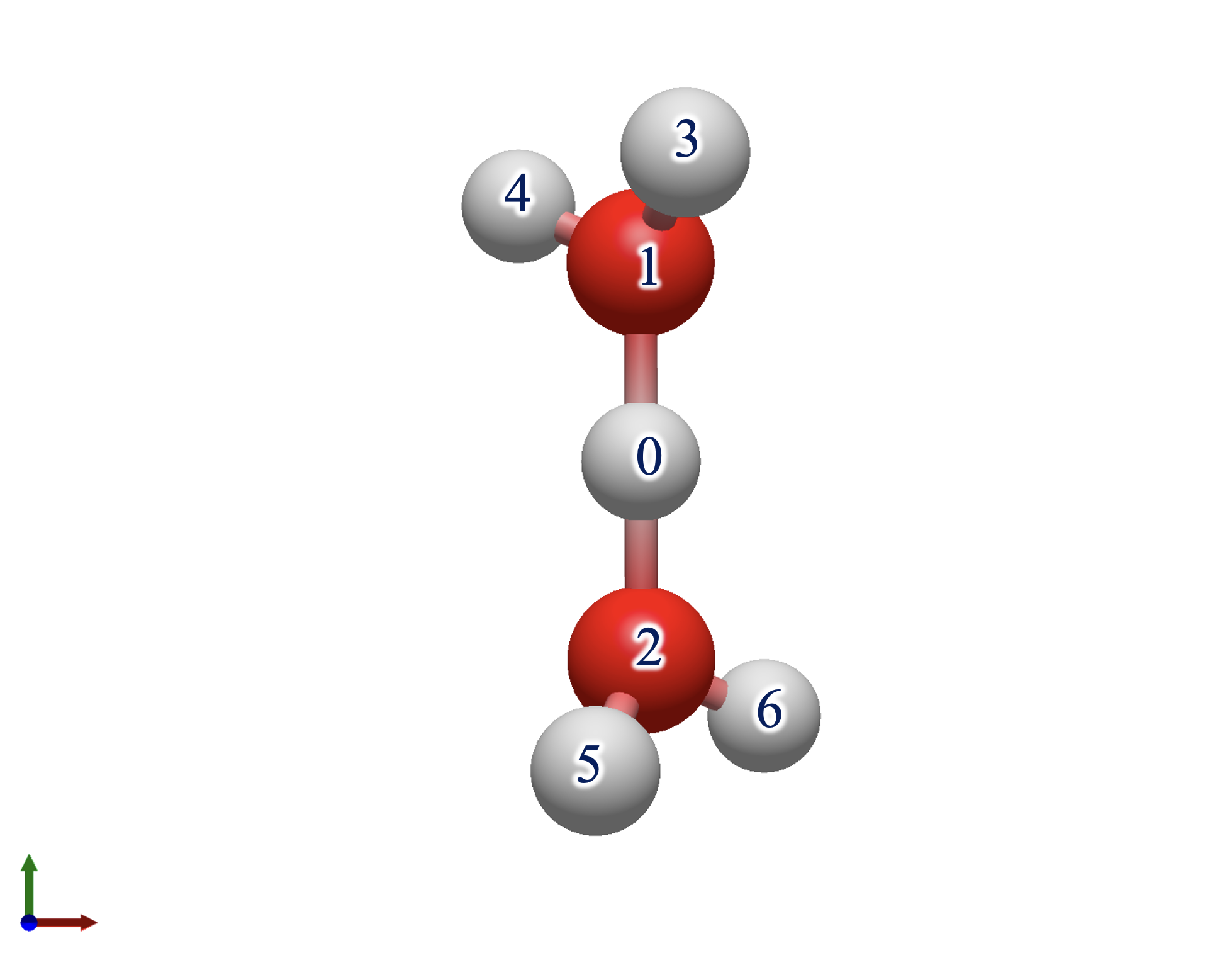}
                \caption{Geometry of the Zundel (H$_{5}$O$_{2}^{+}$) cation, calculated with CNEO-MP2, using the aug-cc-pVTZ electronic basis set, and the PB4-D nuclear basis set for the shared proton. The optimized structure has C$_{2}$ symmetry. This visualization was generated with Avogadro 2.0 \cite{Avogadro}. The blue arrow corresponds to the positive z-axis, the red arrow to the positive x-axis, and the green arrow to the positive y-axis.}
    \label{fig:placeholder}
\end{figure}
\FloatBarrier

\newpage
\begin{table}[H]
\begin{center}
\caption{Internuclear distances (\AA ngstroms), bond angles (degrees), and dihedral angles (degrees) of the C$_{2}$ symmetry H$_{5}$O$_{2}^{+}$ ion, calculated with the aug-cc-pVTZ electronic basis set. Results are presented for the single-component MP2, single-component VPT2-MP2, and multicomponent CNEO-MP2 methods. All calculations used the aug-cc-pVTZ electronic basis set. For the CNEO-MP2 calculations, only the shared hydrogen nucleus was treated quantum mechanically with the PB4-D nuclear basis set. The atom indexing corresponds to the indexing labels used in Figure 5.}
\begin{tabular}{|| c | c c c ||}
\hline
TYPE & MP2 & VPT2-MP2 & CNEO-MP2\\
\hline
O(1)\textendash O(2) & 2.391 & 2.434 & 2.411 \\
O(1)\textendash H(3) & 0.968 & 0.954 & 0.966 \\
O(1)\textendash H(4) & 0.969 & 0.955 & 0.966 \\
O(2)\textendash H(5) & 0.968 & 0.954 & 0.966 \\
O(2)\textendash H(6) & 0.969 & 0.955 & 0.966 \\
H(0)\textendash O(1) & 1.197 & 1.218 & 1.207 \\
H(0)\textendash O(2) & 1.197 & 1.218 & 1.207 \\
$\theta$(O(1)\textendash H(0)\textendash O(2)) & 173.628 & 174.7898 & 173.502 \\
$\theta$(H(3)\textendash O(1)\textendash H(4)) & 108.778 & 109.744 & 108.928 \\
$\theta$(H(5)\textendash O(2)\textendash H(6)) & 108.778 & 109.744 & 108.928 \\
$\theta$(H(3)\textendash O(1)\textendash H(0)) & 118.003 & 120.498 & 118.523 \\
$\theta$(H(4)\textendash O(1)\textendash H(0)) & 116.102 & 118.668 & 116.610 \\
$\theta$(H(5)\textendash O(2)\textendash H(0)) & 118.003 & 120.487 & 118.522 \\
$\theta$(H(6)\textendash O(2)\textendash H(0)) & 116.102 & 118.673 & 116.610 \\
$\tau$(H(3)\textendash O(1)\textendash O(2)\textendash H(5)) & 33.905 & 43.200 & 35.654 \\
$\tau$(H(3)\textendash O(1)\textendash O(2)\textendash H(6)) & -99.774 & -98.840 & -99.573 \\
$\tau$(H(4)\textendash O(1)\textendash O(2)\textendash H(5)) & -99.774 & -98.840 & -99.573 \\
$\tau$(H(4)\textendash O(1)\textendash O(2)\textendash H(6)) & 126.547 & 119.130 & 125.199 \\
\hline
\end{tabular}
\end{center}
\end{table}

The internuclear distances between the oxygen atoms and the shared proton increase with both VPT2-MP2 and CNEO-MP2, as expected.
Subsequently, numerical Hessians were calculated for these structures. For the CNEO-MP2 method, the correlated density was constrained in the energy evaluations. The calculated vibrational frequencies are presented in Table 7, alongside comparisons with numerical frequencies calcualted with MP2 in ORCA 6.0.1, and anharmonically corrected numerical frequencies calculated with VPT2-MP2 in Gaussian 16, all of which were calculated with the aug-cc-pVTZ electronic basis set.

\begin{table}[H]
\begin{center}
\caption{Vibrational frequencies (cm$^{-1}$) of the Zundel cation (H$_{5}$O$_{2}^{+}$), calculated with the aug-cc-pVTZ electronic basis set. Results are presented for single-component MP2, VPT2-MP2, and CNEO-MP2. All calculations used the aug-cc-pVTZ electronic basis set. For the CNEO-MP2 calculations, only the shared proton was treated quantum mechanically with the PB4-D basis set. The geometry was optimized with CNEO-MP2 UCD and the numerical Hessian calculations were performed with both CNEO-MP2 UCD and CNEO-MP2 CCD.}
\begin{tabular}{|| c | c | c c c ||}
\hline
Mode & Type & MP2 & VPT2-MP2 & CNEO-MP2 \\
\hline
1 & free O\textendash H symm. asymm. stretch & 3837.0 & 3658.0 & 3860.9 \\
2 & free O\textendash H asymm. asymm. stretch & 3836.6 & 3655.3 & 3860.3 \\
3 & free O\textendash H symm. symm. stretch & 3740.7 & 3574.5 & 3768.0 \\
4 & free O\textendash H asymm. symm. stretch & 3734.3 & 3568.3 & 3752.0 \\
5 & concerted bend/O\textendash H\textendash O stretch & 1761.1 & 1846.8 & 1795.5 \\
6 & concerted bend/H osc. (z) & 1704.6 & 1636.2 & 1693.6 \\
7 & concerted bend/H osc. (z) & 1546.6 & 1392.5 & 1521.6 \\
8 & concerted bend/H osc. (x) & 1475.0 & 1305.5 & 1450.3 \\
9 & O\textendash H\textendash O stretch/H$_{2}$O wagging & 912.1 & 1507.2 & 1251.1 \\
10 & H$_{2}$O rocking/O\textendash O stretch & 623.4 & 281.4 & 607.8 \\
11 & H$_{2}$O wagging/H osc. (z) & 540.6 & 451.5 & 533.0 \\
12 & H$_{2}$O rocking/H osc. (prim. x) & 533.2 & 467.0 & 517.6 \\
13 & H$_{2}$O wagging/H osc. (z) & 470.2 & 367.5 & 450.5 \\
14 & O\textendash H\textendash O stretch/H$_{2}$O wagging & 388.1 & 475.6 & 421.7 \\
15 & (H$_{2}$O) $\cdot$ H $\cdot$ (H$_{2}$O) torsion & 191.8 & 151.6 & 174.3 \\
\hline
\end{tabular}
\end{center}
\end{table}

In the experimental vibrational spectra of the Zundel ion, two prominent bands occur, one at 1080 cm$^{-1}$  attributed to oscillation of the shared proton,  and one at 1770 cm$^{-1}$ attributed to a concerted bending motion of the system. \cite{Headrick2} For these peaks in particular, single component MP2 is in error by nearly 170 cm$^{-1}$ for the shared proton stretching mode, but only by about 9 cm$^{-1}$ for the bending mode. VPT2-MP2 shifts the frequencies in the right direction (they are blueshifted relative to MP2 towards the experimental frequencies), but significantly overestimates the frequencies, by nearly 430 cm$^{-1}$ for the stretching mode, and by about 76 cm$^{-1}$ for the bending mode. In comparison, CNEO-MP2 also shifts the peaks in the correct direction (i.e. blueshifts away from the single-component MP2 frequencies and towards those of the experimental bands) and while it still overestimates the frequencies, by about 171 cm$^{-1}$  for the stretching mode and nearly 26 cm$^{-1}$  for the bending mode, it performs significantly better than VPT2-MP2 in describing these important modes involving motion of the shared proton. 
While previous path integral molecular dynamics studies investigating nuclear quantum effects upon vibrational frequencies have found that including nuclear quantum effects upon calculated frequencies generally results in redshifting of the calculated frequencies \cite{PIMD3, PIMD4, PIMD1, PIMD2}  (which does still occur for most of the VPT2-MP2 and CNEO-MP2 calculated vibrational frequencies) for the Zundel ion in particular some blueshifting  of frequencies is expected, arising from the large amplitude motion of the proton transfer mode due to the double well nature of the potential energy surface associated with this mode.\cite{Dahms} While both CNEO-MP2 and VPT2-MP2 capture the expected redshifting and the blueshifting for the important shared proton modes, CNEO-MP2 outperforms VPT2-MP2 for these modes, which provides evidence for the potential utility of the CNEO-MP2 method for future applications.


\section{Conclusion}
In this study, a new multicomponent MP2 method was introduced in combination with the CNEO formalism called CNEO-MP2. When comparing the bond lengths calculated with CNEO-MP2 to those of the single-component MP2 and single-component VPT2-MP2, the CNEO-MP2 method, like VPT2-MP2, correctly produces longer bond lengths seen when nuclear quantum effects are included in computational chemistry calculations. It is also important to note that for all of the single-component MP2 results, there is no effect upon bond lengths upon substitution of hydrogen with deuterium. For the CNEO-MP2 results however, as with VPT2-MP2, shorter bond lengths are obtained for the deuterium substituted isotopes. 

Thus the CNEO-MP2 method correctly describes the changes in molecular geometries due to vibrational effects, including correctly capturing the isotopic effects. This is accomplished directly with CNEO-MP2 by treating the nuclei as quantum mechanical particles. Importantly, as a result of this direct treatment of nuclear quantum effects, these effects are able to be captured in a single calculation or geometry optimization. 

The CNEO-MP2 method performs well for the calculation of vibrationally averaged molecular properties, and the Lagrangian formulation presented provides a good foundation for the development of further wavefunction approaches within the CNEO framework. Building upon this foundation, as well as previous developments of multicomponent NEO coupled cluster methods, \cite{Pav, NEO246, NEO259, NEO298} our future plans include the development of a multicomponent coupled cluster method within the CNEO framework.


\section{Conflict of Interest}
The authors have no conflicts to disclose.

\section{Author Contributions}
\textbf{Gabrielle B. Tucker}: methodology (lead), software (lead), validation (lead), writing/original draft preparation (lead), writing/review and editing (equal). 

\textbf{Kurt R. Brorsen}: conceptualization (lead), methodology (supporting), supervision (lead), writing/original draft preparation (supporting), writing/review and editing (equal).

\section{Data Availability Statement}
The data that supports the findings of this study are available within the article and its supplementary material.

\section{Supplementary Materials}
The full data for the CNEO-MP2-CCD and CNEO-MP2-UCD calculations and the geometries used for the proton affinity calculations are available.

\section{Acknowledgments}
This material is based upon work supported by the National Science Foundation under Grant No. 2418760.

\end{document}